\documentclass[12pt]{article}
\usepackage{graphicx}

\newcommand\pubnumber{}
\newcommand\pubdate{\today}

\def\support{\footnote{on behalf of the Belle collaboration.}}

\def\Title#1{\begin{center} {\Large #1 } \end{center}}
\def\Author#1{\begin{center}{ \sc #1} \end{center}}
\def\Address#1{\begin{center}{ \it #1} \end{center}}

\newcommand\pubblock{\rightline{\begin{tabular}{l} \pubnumber\\
         \pubdate  \end{tabular}}}
\newenvironment{Abstract}{\begin{quotation}  }{\end{quotation}}
\newenvironment{Presented}{\begin{quotation} \begin{center} 
             PRESENTED AT\end{center}\bigskip 
      \begin{center}\begin{large}}{\end{large}\end{center} \end{quotation}}
\def\Acknowledgements{\bigskip  \bigskip \begin{center} \begin{large}
             \bf ACKNOWLEDGEMENTS \end{large}\end{center}}

\def\beq{\begin{equation}}
\def\eeq#1{\label{#1}\end{equation}}
\def\eeqn{\end{equation}}

\def\beqa{\begin{eqnarray}}
\def\eeqa#1{\label{#1}\end{eqnarray}}
\def\eeqan{\end{eqnarray}}

\let\bar=\overbar

\def\etal{{\it et al.}}

\def\Dslash{\not{\hbox{\kern-4pt $D$}}}
\def\dslash{\not{\hbox{\kern-2pt $\del$}}}

\def\msb{{\bar{\ssstyle M \kern -1pt S}}}

\begin{document}
\begin{titlepage}
\pubblock

\vfill
\Title{New Belle results on $D^0 - \bar{D}^0$ mixing}
\vfill
\Author{Marko Stari\v c\support}
\Address{J. Stefan Institute, Jamova 39, 1000 Ljubljana, Slovenia}
\vfill
\begin{Abstract}
  We report the new measurement of $D^0 - \bar{D}^0$ mixing in decays to $K^+K^-$ and
  $\pi^+\pi^-$ final states that is based on the total Belle data sample of 976~fb$^{-1}$.
  The preliminary results are $y_{CP}=(1.11 \pm 0.22 \pm 0.11)\%$ and
  $A_{\Gamma}=(-0.03 \pm 0.20 \pm 0.08)\%$.
\end{Abstract}
\vfill
\begin{Presented}
  The 5$^{\rm th}$ International Workshop on Charm Physics (Charm~2012)\\
  14-17 May 2012, Honolulu, Hawaii 96822.
\end{Presented}
\vfill
\end{titlepage}
\def\thefootnote{\fnsymbol{footnote}}
\setcounter{footnote}{0}

\section{Introduction}

Mixing of neutral mesons occurs when the flavor eigenstates differ from the physical 
mass eigenstates of the meson-antimeson system. In case of $D^0$ mesons mass eigenstates
are expressed as $|D^0_{1,2}\rangle = p|D^0\rangle\pm q|\bar{D^0}\rangle$, 
with $p^2+q^2=1$. If $p=q$ the two mass eigenstates are CP-even and CP-odd;
otherwise $CP$ is violated. Charm mixing is 
characterized by two parameters: $x = \Delta m/\Gamma$ and $y = \Delta \Gamma/2\Gamma$,
where $\Delta m$ and $\Delta \Gamma$ are the mass and decay width differences of mass
eigenstates, respectively, and $\Gamma$ is the average $D^0$ decay width.

Mixing in $D^0$ decays to $CP$ eigenstates, such as $D^0 \to K^+K^-$, manifests in 
a lifetime that differs from the lifetime of decays to flavor eigenstates, such as
$D^0 \to K^-\pi^+$. The quantity
\begin{equation}
  y_{CP} = \frac{\tau(K^-\pi^+)}{\tau(K^+K^-)} - 1
\end{equation} 
is equal to the mixing parameter $y$ if $CP$ is conserved. If $CP$ is violated 
the lifetimes of $D^0$ and $\bar{D}^0$ decaying to the same $CP$ eigenstate 
also differ and the lifetime asymmetry, defined as
\begin{equation}
  A_{\Gamma} = \frac{\tau(\bar{D}^0 \to K^-K^+)-\tau(D^0 \to K^+K^-)}
  {\tau(\bar{D}^0 \to K^-K^+)+\tau(D^0 \to K^+K^-)}
\end{equation}
becomes non-zero. The quantities $y_{CP}$ and $A_{\Gamma}$ are connected to the
mixing parameters $x$ and $y$ by~\cite{Bergmann}
$y_{CP} = y \cos{\phi} - \frac{1}{2} A_M x \sin{\phi}$ and
$A_\Gamma = \frac{1}{2} A_M y \cos{\phi} - x \sin{\phi}$,
where $\phi = \arg(q/p)$ and $A_M = 1 - |q/p|^2$.

First evidence for $D^0 - \bar{D}^0$ mixing was obtained in 2007 by Belle~\cite{Belle}
and by BaBar~\cite{Babar} in two different decay modes. We report here an update of
our first-evidence measurement in $D^0 \to K^+K^-, \pi^+\pi^-$ decays using almost
twice larger data set.

\section{Event selection}

 The decays $D^0 \to K^+K^-, \pi^+\pi^-$ and $D^0 \to K^-\pi^+$ are reconstructed in
the decay chain $D^{*+} \to D^0\pi^+$ in order to suppress background and to tag the 
$D^0$ flavor at production. To reject $D^{*+}$ candidates coming from $B$ decays, we
require that the $D^{*+}$ momentum measured in the center-of-mass system (CMS) 
of the $e^+e^-$ collisions be greater that 2.5~GeV/c; 
for a fraction of data taken at $\Upsilon(5S)$ we increase this threshold to 3.1~GeV/c. 

Besides using our standard kaon and pion selection criteria, 
based on $dE/dx$, threshold aerogel
Cherenkov counters and time-of-flight, and vertex fits, we select $D^0$ candidates
using two kinematic variables: the invariant mass of the $D^0$ decay products $M$
and the energy released in the $D^{*+}$ decay $q=(M_{D^*}-M-m_\pi)c^2$.
The proper  decay time of the
$D^0$ candidate is calculated from the projection of the vector
joining the two vertices, $\vec{L}$, onto the $D^0$ momentum vector: 
$t=m_{D^0}\vec{L}\cdot\vec{p}/p^2$, where
$m_{D^0}$ is the nominal $D^0$ mass~\cite{PDG}. The decay time uncertainty 
$\sigma_t$ is evaluated event-by-event from the error
matrices of the production and decay vertices.

The sample of events for the lifetime measurements
is selected using $\Delta M$, $\Delta q$ and $\sigma_t$. These selection criteria
are optimized on Monte Carlo (MC) simulation by minimizing the statistical error
on $y_{CP}$. Background is estimated from sidebands in $M$; the sideband position 
has also been optimized in order to minimize the systematic uncertainty. 
The yields of selected
events are $242 \times 10^3$ ($K^+K^-$), $114 \times 10^3$ ($\pi^+\pi^-$) and
$2.61 \times 10^6$ ($K^-\pi^+$), with signal purities of 98.0\%, 92.9\% and 99.7\%,
respectively.

\section{Lifetime fit}

By studying the proper decay time distribution of $D^0 \to K^-\pi^+$ decays
we observe a significant dependence of its mean value on $\cos \theta^*$, 
where $\theta^*$ is the $D^0$ CMS polar angle. Using MC simulation we find that
the generated proper decay time distribution of selected events
agrees well in each $\cos \theta^*$ bin with an exponential distribution and that 
the lifetime is consistent with the generated value. 
The observed dependence is thus due to the resolution 
function offset that depends on $\cos \theta^*$. Therefore, to reduce
systematic uncertainties arising from the resolution function parameterization
the measurement is performed in bins of $\cos \theta^*$;
an additional requirement $|\cos \theta^*|<0.9$ is imposed to suppress the events 
with the largest offsets (about 1\% of events).

The fitting procedure in each $\cos \theta^*$ bin is similar to the one used in
our previous measurement~\cite{Belle}, where we performed a binned simultaneous 
fit to $KK$, $K\pi$ and $\pi\pi$ samples. 
The resolution function is constructed similarly, 
from normalized distribution of $\sigma_t$ using a double or triple Gaussian PDF for
each $\sigma_t$ bin. The widths $\sigma_k^{\rm pull}$ and fractions $w_k$ of these
Gaussian distributions are obtained from fits to the MC distribution of pulls
$(t-t_{\rm gen})/\sigma_t$. The procedure is repeated for each $\cos \theta^*$ bin.
The parameterization reads:
\begin{equation}
  R(t)=\sum_{i=1}^{n_{\rm bin}} f_i \sum_{k=1}^{n_g} w_k G(t;\mu_i, \sigma_{ik}),
\end{equation}
where $G(t;\mu_i, \sigma_{ik})$ represents a Gaussian distribution of mean $\mu_i$ and
width $\sigma_{ik}$, and $f_i$ is the fraction of events in the $i$-th bin of the
$\sigma_t$ distribution. The mean and width are parameterized as:
\begin{equation}
  \sigma_{ik}= s_k \sigma_k^{\rm pull}\sigma_i~~~~~~~~~~~
  \mu_i=t_0 + a (\sigma_i - \sum_{j=1}^n f_j \sigma_j)
\end{equation}
where $s_k,~k=1,...,n_g$ are the width scaling factors for each of the $n_g$ Gaussian's,
$t_0$ is the resolution function offset, and $a$ is a parameter to model a possible
asymmetry of the resolution function. The parameters $s_k$, $t_0$ and $a$ are
free parameters in the fit.

The proper decay time distribution is parameterized with
\begin{equation}
  f(t)=\frac{N}{\tau}\int e^{-t'/\tau} R(t-t') dt' + B(t),
\end{equation}
with the following free parameters: $N$, $\tau$, $s_k$, $t_0$ and $a$. 
A sideband subtracted $\sigma_t$ distribution is used to construct $R(t)$.
The term $B(t)$ describes background and is fixed with a fit to the sideband 
distribution. 

Background is parameterized as a sum of two lifetime components, 
a component with zero lifetime and a component with an effective lifetime $\tau_b$:
\begin{equation}
  B(t)=N_b \int [f \delta(t') + (1-f) \frac{1}{\tau_b}e^{-t'/\tau_b}]R_b(t-t') dt'
\end{equation}
The background resolution function $R_b(t)$ is assumed to be symmetric 
($a \equiv 0$) and is
composed of three Gaussian's with $s_3 = s_2$. The fraction $f$ of zero-lifetime 
component is found to be $\cos \theta^*$ dependent; its value is fixed in each bin 
using MC simulation. 
The parameters $t_0$, $s_1$, $s_2$ and $\tau_b$ are determined from
a fit to sideband distribution summed over $\cos \theta^*$ bins. However, the shape
in individual bins remains bin-dependent since $\sigma_t$ distribution, $f$ and
$N_b$ depend on $\cos \theta^*$.

To extract $y_{CP}$ and $A_\Gamma$ the decay modes are fitted simultaneously in
each $\cos \theta^*$ bin using a binned maximum likelihood fit. The parameters
shared between decay modes are $y_{CP}$ and $A_\Gamma$ ($KK$ and $\pi\pi$),
$t_0$ and $a$ (all decay modes) and parameters $s_1$, $s_2$ and $s_3$ 
up to an overall scaling factor. 
The fit has been tested with the generic MC simulation equivalent to
six times the data statistics. The fitted $y_{CP}$ and $A_\Gamma$ are found to
be consistent with the input zero value and the fitted $K\pi$ lifetime $\tau$ is found
to be consistent with the generated value. A linearity test performed with MC simulated
events re-weighted to reflect different $y_{CP}$ shows no bias.

\section{Results}

The experimental data were taken with two different silicon vertex detector (SVD)
configurations: for the first 
153~fb$^{-1}$ a 3-layer SVD was used, while for the rest of the data a 4-layer SVD
was used. We treat both running periods separately, since the resolution function
differs. 

\begin{figure}[htb]
  \centering
  \includegraphics[width=.8\textwidth]{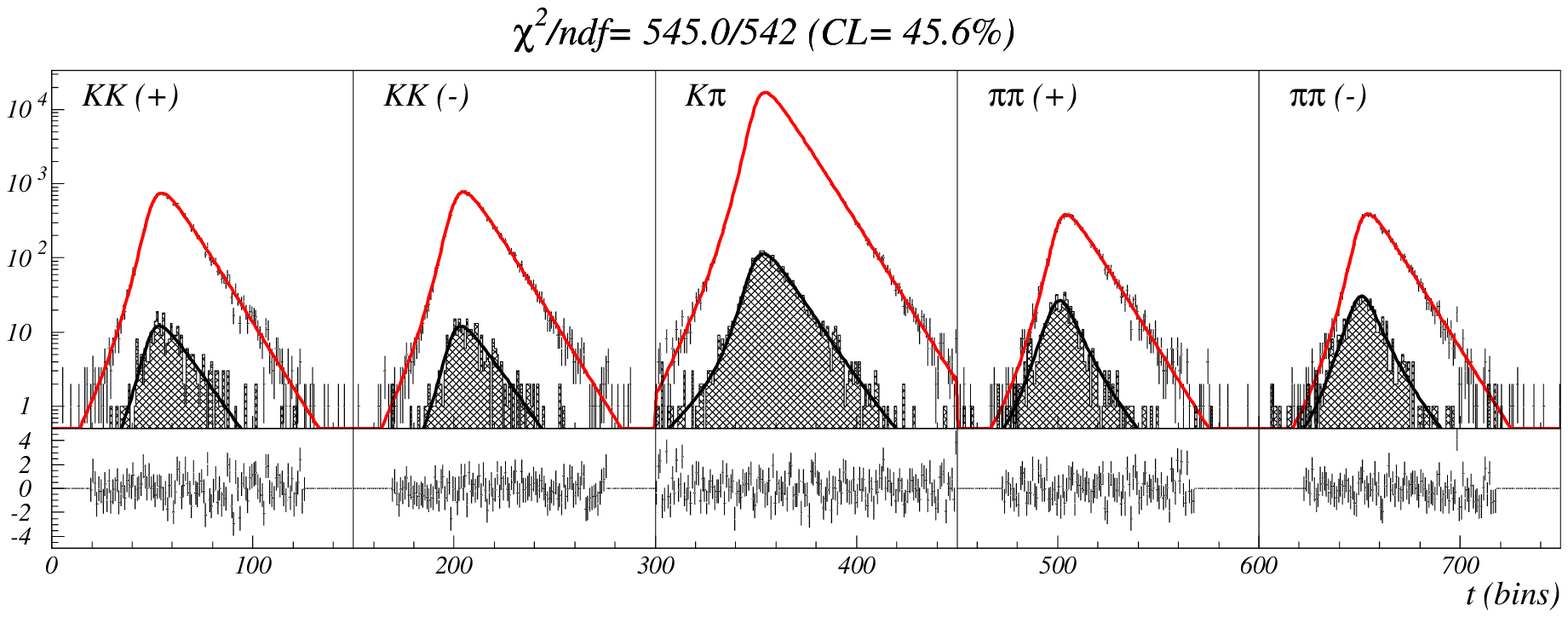}
  \includegraphics[width=.8\textwidth]{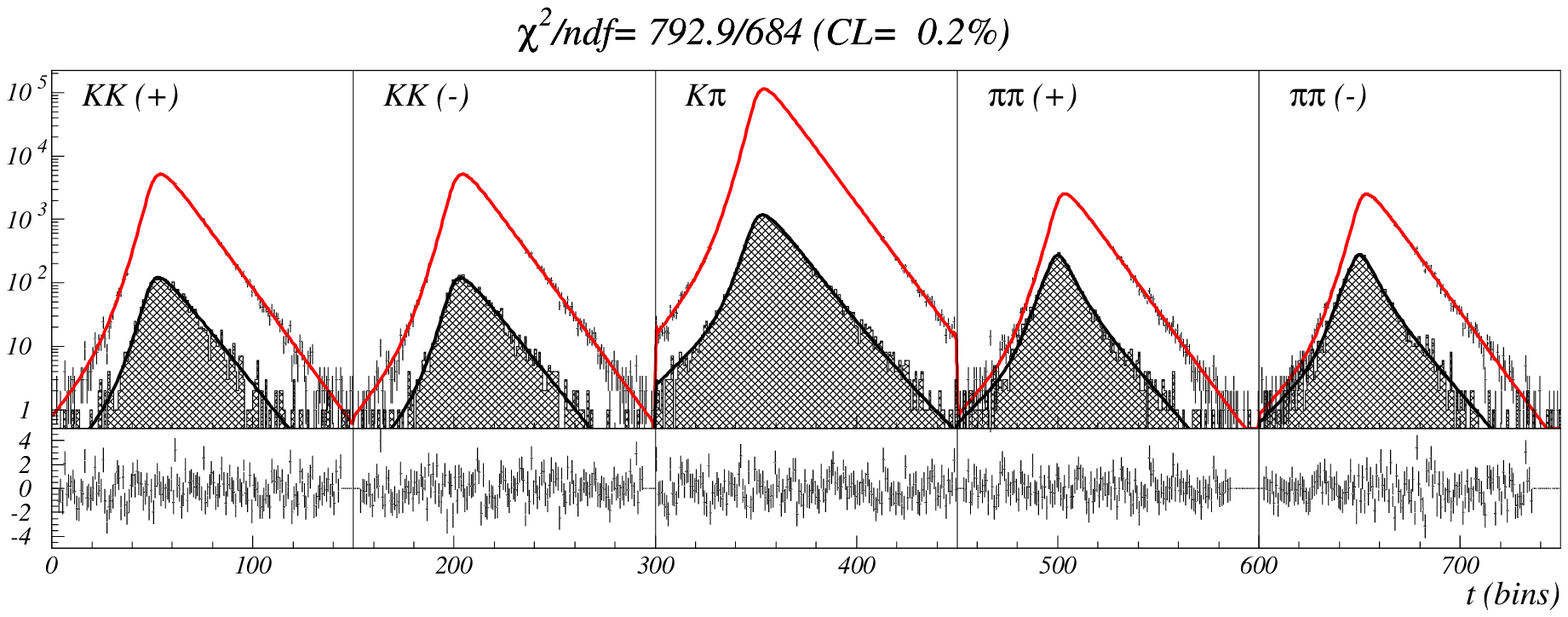}
  \caption{Simultaneous fit of SVD1 (top) and SVD2 (bottom) proper decay time
  distributions. The plots show a sum of distributions 
  and fitted functions (in red) over $\cos \theta^*$ bins; the residuals are 
  plotted beneath. The signal-region distributions are shown as error bars and the
  sideband-region distributions as hatched histograms. The ``(+)'' and the ``(-)'' 
  denote the charge of the tagging slow pion.}
  \label{fits.eps}
\end{figure}

The proper decay time distributions are fitted simultaneously as discussed in the
previous section. The results of the fits are shown in Fig.~\ref{fits.eps}.  
Fit confidence levels (CL) are above 5\% (except one with CL=3.3\%)
and are distributed uniformly~\footnote{We use Pearson's definition of $\chi^2$ and 
take only the bins with the fitted function greater than one.}. 
The plots in Fig.~\ref{fits.eps} are obtained by summing the histograms 
and functions in all fitted $\cos \theta^*$ bins. The residuals
show no significant structure. The normalized Pearson's $\chi^2$ are 1.01 (SVD1) and
1.16 (SVD2). 

\begin{figure}[htb]
  \centering
  \includegraphics[width=.8\textwidth]{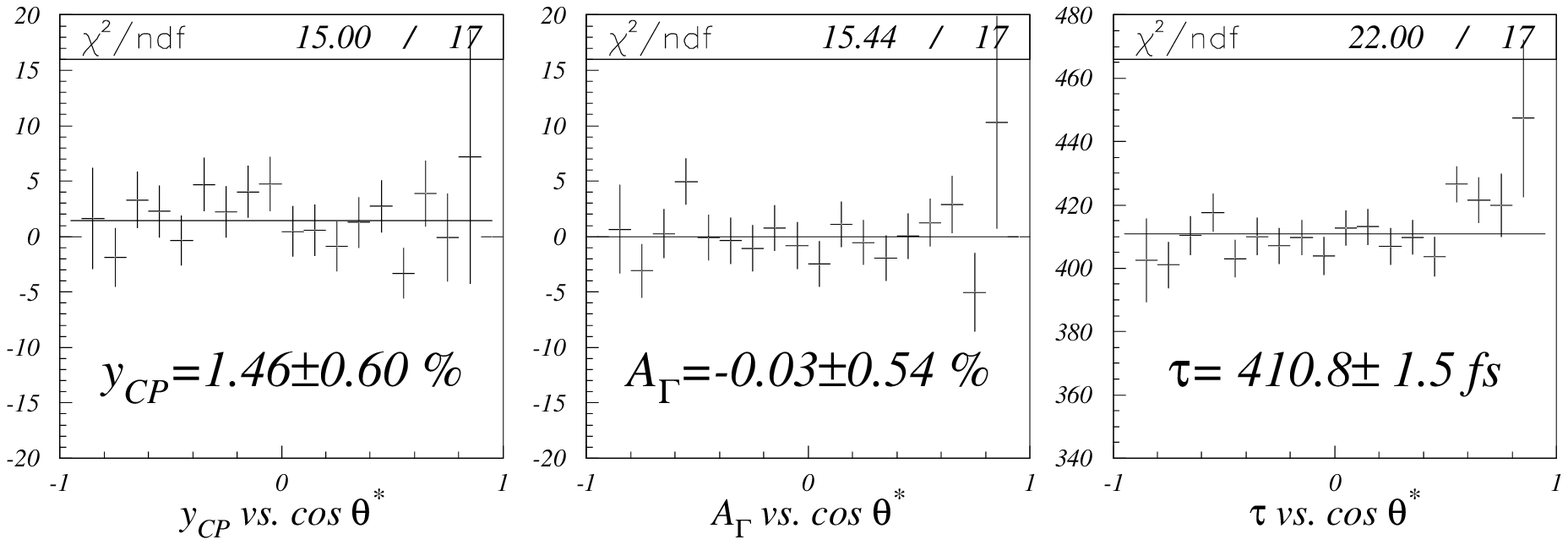}
  \includegraphics[width=.8\textwidth]{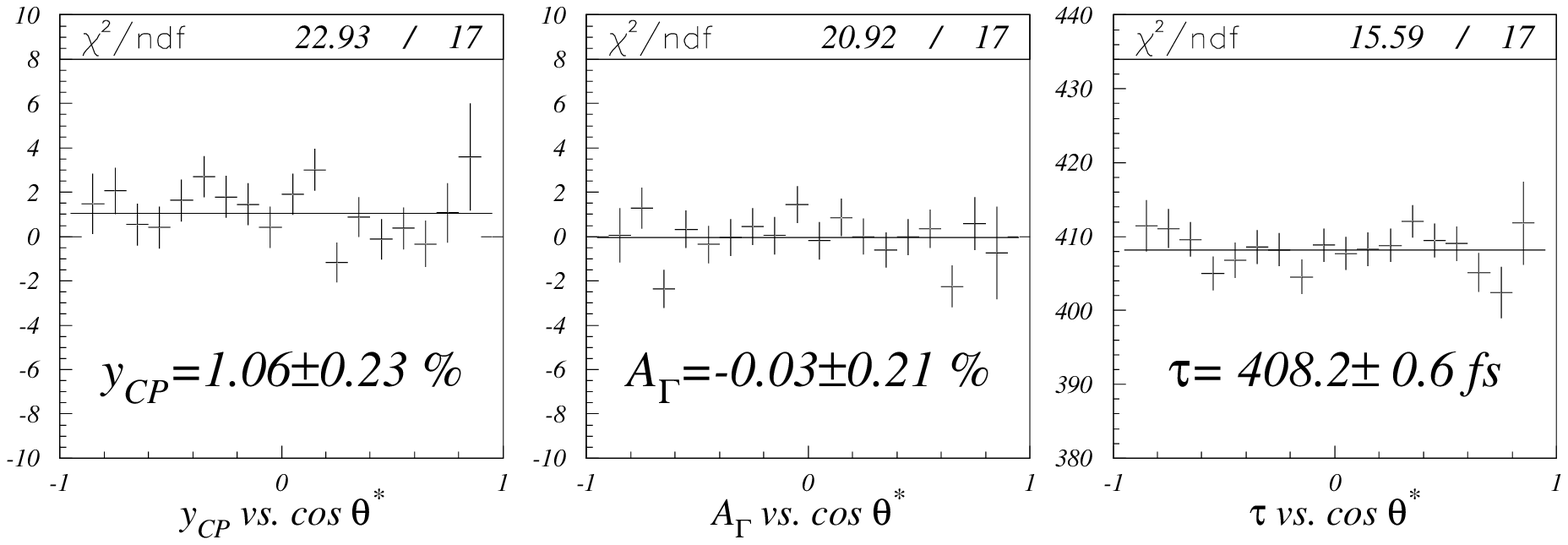}
  \caption{Results of simultaneous fits in bins of $\cos \theta^*$ 
    (points with error bars) for SVD1 (top) and SVD2 (bottom). 
    The horizontal line in each plot is a least square fit of constant to data points.}
  \label{results.eps}
\end{figure}

Fig.~\ref{results.eps} shows the results of the fits in bins of $\cos \theta^*$ for
$y_{CP}$, $A_\Gamma$ and the $D^0 \to K^-\pi^+$ lifetime $\tau$. The average is
obtained by a least square fit to a constant. We find $y_{CP} = (1.11\pm 0.22)\%$,
$A_\Gamma = (-0.03 \pm 0.20)\%$ and $\tau = (408.46\pm 0.54)~{\rm fs}$, where the
errors are statistical only. The results for individual running periods are consistent
with each other. The measured $D^0$ lifetime is also consistent with the current
world average~\cite{PDG}. 

\section{Systematics}

The systematic uncertainties are summarized in Table~\ref{syst.tab}. The largest
contribution is found to arise from possible SVD misalignments. 
The impact of misalignments has
been extensively studied using a special signal MC simulation with different local 
and global SVD misalignments. We find that the SVD misalignment affects
the resolution function considerably, especially its offset $t_0$, and it can explain 
the differences seen between MC and data. However, the effect on the resolution function 
is very similar for $KK$, $K\pi$ and $\pi\pi$ decay modes, and thus mainly cancels 
when relative lifetime measurements such as $y_{CP}$ and $A_\Gamma$ are performed.

Other sources of systematics arise from possible impact of selection criteria to 
the acceptance, from the position of mass window $\Delta M$, 
from background parameterization 
and from background statistical fluctuations, from resolution function parameterization 
and from binning in $t$ and $\cos \theta^*$.
The total systematic uncertainties are estimated to 0.11\% for $y_{CP}$ and
0.08\% for $A_\Gamma$.

\begin{table}[t]
  \caption{Systematic uncertainties}
  \label{syst.tab}
  \begin{center}
    \begin{tabular}{|l|cc|}
      \hline
      source & $\Delta y_{CP}$ (\%) & $\Delta A_\Gamma$ (\%) \\
      \hline
      acceptance           & 0.050 & 0.044 \\
      SVD misalignments    & 0.060 & 0.041 \\
      mass window position & 0.007 & 0.009 \\
      background           & 0.059 & 0.050 \\
      resolution function  & 0.030 & 0.002 \\
      binning              & 0.021 & 0.010 \\
      \hline
      sum in quadrature    & 0.11 & 0.08 \\
      \hline
    \end{tabular}
  \end{center}
\end{table}

\section{Summary}

With the full Belle data set of 976~fb$^{-1}$, we obtain the preliminary results
\begin{eqnarray}
  y_{CP} &=& (+1.11 \pm 0.22 \pm 0.11)\%, \\
  A_\Gamma &=& (-0.03 \pm 0.20 \pm 0.08)\%,
\end{eqnarray}
where the first error is statistical and the second is systematical. 
The significance of $y_{CP} \neq 0$ is 4.5$\sigma$ when both errors are combined 
in quadrature,
and 5.1$\sigma$ if only the statistical error is considered. The result for $A_\Gamma$ is
consistent with no indirect $CP$ violation. Both results are in good agreement with
our previous measurement~\cite{Belle} as well as with the BaBar measurements in the 
same decay modes~\cite{Babar-1, Babar-2}.

\Acknowledgements
We thank the KEKB group for the excellent operation of the
accelerator; the KEK cryogenics group for the efficient
operation of the solenoid; and the KEK computer group,
the National Institute of Informatics, and the 
PNNL/EMSL computing group for valuable computing
and SINET4 network support.  We acknowledge support from
the Ministry of Education, Culture, Sports, Science, and
Technology (MEXT) of Japan, the Japan Society for the 
Promotion of Science (JSPS), and the Tau-Lepton Physics 
Research Center of Nagoya University; 
the Australian Research Council and the Australian 
Department of Industry, Innovation, Science and Research;
the National Natural Science Foundation of China under
contract No.~10575109, 10775142, 10875115 and 10825524; 
the Ministry of Education, Youth and Sports of the Czech 
Republic under contract No.~LA10033 and MSM0021620859;
the Department of Science and Technology of India; 
the Istituto Nazionale di Fisica Nucleare of Italy; 
he BK21 and WCU program of the Ministry Education Science and
Technology, National Research Foundation of Korea Grant No.\ 
2010-0021174, 2011-0029457, 2012-0008143, 2012R1A1A2008330,
BRL program under NRF Grant No. KRF-2011-0020333,
and GSDC of the Korea Institute of Science and Technology Information;
the Polish Ministry of Science and Higher Education and 
the National Science Center;
the Ministry of Education and Science of the Russian
Federation and the Russian Federal Agency for Atomic Energy;
the Slovenian Research Agency;
the Basque Foundation for Science (IKERBASQUE) and the UPV/EHU under 
program UFI 11/55;
the Swiss National Science Foundation; the National Science Council
and the Ministry of Education of Taiwan; and the U.S.\
Department of Energy and the National Science Foundation.
This work is supported by a Grant-in-Aid from MEXT for 
Science Research in a Priority Area (``New Development of 
Flavor Physics''), and from JSPS for Creative Scientific 
Research (``Evolution of Tau-lepton Physics'').


\begin{thebibliography}{99}

\bibitem{Bergmann} S. Bergmann \etal, Phys. Lett. B {\bf 486}, 418 (2000)
\bibitem{Belle} M. Stari\v c \etal (Belle Collaboration), 
  Phys. Rev. Lett. {\bf 98}, 211803 (2007)
\bibitem{Babar} B. Aubert \etal (BaBar Collaboration), 
  Phys. Rev. Lett. {\bf 98}, 211802 (2007)
\bibitem{PDG} K. Nakamura \etal (Particle Data Group), J. Phys. G: Nucl. Part. Phys.
  {\bf 37} (2010) 075021
\bibitem{Babar-1} B. Aubert \etal (BaBar Collaboration), 
  Phys. Rev. {\bf D78} (2008) 011105
\bibitem{Babar-2} B. Aubert \etal (BaBar Collaboration), 
  Phys. Rev. {\bf D80} (2009) 071103
\end{thebibliography}
\end{document}